\documentclass[12pt]{article}
\usepackage{graphicx}
\usepackage{amsmath, amssymb}

\newcommand{\half}{{\frac{1}{2}}}
\newcommand{\ket}[1]{\,\left|\,{#1}\right\rangle}
\newcommand{\mbf}[1]{\mathbf{#1}}

\renewcommand{\bar}[1]{\overline{#1}}

\def\Dslash{\raise.15ex\hbox{/}\kern-.7em D}
\def\Pslash{\raise.15ex\hbox{/}\kern-.7em P}

\begin {document}
\begin{flushright}
{\small
SLAC--PUB--12804\\
September 2007}
\end{flushright}

\vspace{20pt}

\centerline{\LARGE \bf AdS/CFT and Exclusive Processes in QCD}

\vspace{15pt}

\centerline{\bf {
Stanley J. Brodsky\footnote{Electronic address:
sjbth@slac.stanford.edu}$^{a}$ and
Guy F. de T\'eramond\footnote{Electronic address:
gdt@asterix.crnet.cr}$^{b,a}$}}

\vspace{10pt}

{\centerline {$^{a}${Stanford Linear Accelerator Center, 
Stanford University, Stanford, CA 94309, USA}}

\vspace{2pt}

{\centerline {$^{b}${Universidad de Costa Rica, San Jos\'e, Costa Rica}}

 \vspace{15pt}

\begin{abstract}

The AdS/CFT correspondence between string theory in AdS space and
conformal field theories in physical space-time leads to an
analytic, semi-classical model for strongly-coupled QCD which has
scale invariance and dimensional counting at short distances and
color confinement at large distances. One can use holography to map
the amplitude describing the hadronic state in the fifth dimension
of Anti-de Sitter space $\rm{AdS}_5$  to the light-front
wavefunctions of hadrons in physical space-time,  thus providing a
relativistic description of hadrons in QCD at the amplitude level.
In particular, we show that there is an exact correspondence between
the fifth-dimensional coordinate of AdS space $z$ and a specific
impact variable $\zeta$ which measures the separation of the quark
and gluonic constituents within the hadron in ordinary space-time.
New relativistic light-front equations in ordinary space-time can
then be derived which reproduce the results obtained using the
5-dimensional theory.  The effective light-front equations possess
elegant algebraic structures and integrability properties. This
connection between the AdS and the light-front representations
allows one to compute the analytic form of the frame-independent
light-front wavefunctions, the fundamental entities which encode
hadron properties and allow the computation of decay constants, form
factors, deeply virtual Compton scattering, exclusive heavy hadron
decays  and other exclusive scattering amplitudes.  As specific
examples we compute the pion coupling constant $f_\pi$ and study the
behavior of the pion form factor $F_\pi(q^2)$ in the space and
time-like regions. We also determine the Dirac nucleon form factors
$F_1^p(q^2)$ and $F_1^n(q^2)$ in the space-like region.

\end{abstract}

\vspace{2pt}

\begin{center}
{\it Invited talk presented at  \\
Workshop on Exclusive Reactions at High Momentum Transfer\\
21-24 May 2007  \\
Newport News, Virginia\\
 }
\end{center}

\vfill

\newpage

\parindent=1.5pc
\baselineskip=16pt

\setcounter{footnote}{0}

\section{Exclusive Processes in QCD}

Exclusive processes provide essential tests of the quark and gluon
structure of hadrons at the amplitude level and a window into the
form of bound-state wavefunctions of QCD.  For example, hadron form
factors and the generalized distributions which control deeply
virtual Compton scattering are expressible in terms of the
light-front wavefunctions $\psi^H_n(x_i, \mbf{k}_{\perp i},
\lambda_i)$, the bound-state eigensolutions of the QCD light-front
Hamiltonian.

Light-front wavefunctions are the relativistic generalizations of
the familiar Schr\"o-dinger wavefunctions of atomic physics, but they
are determined at fixed light-front time $\tau= t +z/c$---the
``front form" advocated by Dirac~\cite{Dirac:1949cp}---rather than
at fixed ordinary time $t$.   The  light-front wavefunctions  of a
hadron are independent of the momentum of the hadron, and they are
thus boost invariant; Wigner transformations and Melosh rotations
are not required. The light-front formalism for gauge theories in
light-cone gauge is particularly useful in that there are no ghosts
and one has a direct physical interpretation of  orbital angular
momentum.

Given the light-front wavefunctions $\psi_{n/h}$ one can
compute a large range of hadron
observables. For example, the valence and sea quark and gluon
distributions which are measured in deep inelastic lepton scattering
are defined from the squares of the LFWFS summed over all Fock
states $n$. Form factors, exclusive weak transition
amplitudes~\cite{Brodsky:1998hn} such as $B\to \ell \nu \pi$. and
the generalized parton distributions~\cite{Brodsky:2000xy} measured
in deeply virtual Compton scattering  $\gamma^* p \to \gamma p$ are (assuming the ``handbag"
approximation) overlaps of the initial and final LFWFS with $n
=n^\prime$ and $n =n^\prime+2$. The gauge-invariant distribution
amplitude $\phi_H(x_i,Q)$ defined from the integral over the
transverse momenta $\mbf{k}^2_{\perp i} \le Q^2$ of the valence
(smallest $n$) Fock state provides a fundamental measure of the
hadron at the amplitude level~\cite{Lepage:1979zb, Efremov:1979qk};
they  are the nonperturbative input to the factorized form of hard
exclusive amplitudes and exclusive heavy hadron decays in
perturbative QCD. The resulting distributions obey the DGLAP and
ERBL evolution equations as a function of the maximal invariant
mass, thus providing a physical factorization
scheme~\cite{Lepage:1980fj}. In each case, the derived quantities
satisfy the appropriate operator product expansions, sum rules, and
evolution equations. At large $x$ where the struck quark is
far-off shell, DGLAP evolution is quenched~\cite{Brodsky:1979qm}, so
that the fall-off of the DIS cross sections in $Q^2$ satisfies Bloom-Gilman
inclusive-exclusive duality at fixed $W^2.$

One of the most significant theoretical advances in recent years has
been the application of the AdS/CFT
correspondence~\cite{Maldacena:1997re} between string theories
defined in 5-dimensional Anti--de Sitter (AdS) space-time and
conformal field theories in physical space-time.  QCD is not itself
a conformal theory; however there are indications,
both from theory~\cite{vonSmekal:1997is, Furui:2006py}
and phenomenology~\cite{Brodsky:2002nb,Deur:2005cf},
that the QCD coupling is slowly varying at small
momentum transfer. In addition, one can argue that if the gluon has a maximum wavelength or an effective mass~\cite{Cornwall:1981zr} due to confinement, that gluonic vacuum polarization corrections and the $\beta$-function must vanish in the infrared.

If there is a conformal window where the QCD coupling is
large and approximately constant and quark masses can be neglected, then
QCD resembles a conformal theory, thus motivating the application of
AdS/CFT to QCD. Thus, even though QCD is not conformally invariant,
one can use the mathematical representation of the conformal group
in five-dimensional anti-de Sitter space to construct an analytic
first approximation to the theory.  The  resulting AdS/QCD model
gives accurate predictions for hadron spectroscopy and a description
of the quark structure of mesons and baryons which has scale
invariance and dimensional counting at short distances, together
with color confinement at large distances.

As we shall discuss, one can use holography to map the amplitude
$\Phi(z)$ describing the hadronic state in the fifth dimension of
Anti-de Sitter space $\rm{AdS}_5$  to the light-front wavefunctions
$\psi_{n/h}$ of hadrons in physical
space-time~\cite{Brodsky:2006uqa},  thus providing a relativistic
description of hadrons in QCD at the amplitude level. In fact, there
is an exact correspondence between the fifth-dimensional coordinate
of anti-de Sitter space $z$ and a specific impact variable $\zeta$
in the light-front formalism which measures the physical separation
of the constituents within the hadron. One can  derive this
correspondence by noticing that the mapping of $z \to \zeta$
transforms the expression for the form factors in AdS/CFT to the
exact QCD Drell-Yan-West expression in terms of light-front
wavefunctions. The amplitude $\Phi(z)$ describing  the hadronic
state in $\rm{AdS}_5$ can then be precisely mapped to the
light-front wavefunctions $\psi_{n/h}$ of hadrons in physical
space-time~\cite{Brodsky:2006uqa}. One thus obtains not only an
accurate description of the hadron spectrum, but also a
simple but realistic model of the valence light-front wavefunctions
of mesons, baryons, and glueballs. Thus there is a remarkable
mapping between the AdS description of hadrons and the Hamiltonian
formulation of QCD in physical space-time quantized at fixed light
front time $\tau = t + z/c.$

The AdS/QCD correspondence is particularly relevant for the description
of hadro-nic form factors,  since it incorporates the connection between the twist of the hadron to the fall-off of its current matrix elements, as well as  essential aspects of vector meson dominance.
It also provides a convenient framework for analytically continuing the space-like results to the time-like region. Recent applications to the form factors of
mesons and nucleons
~\cite{deTeramond:2006xb, Grigoryan:2007vg, Grigoryan:2007my, Brodsky:2007pt, Brodsky:2007hb, Huang:2007uu, Kwee:2007dd}
have followed from the pioneering work
of refs.~\cite{Polchinski:2001ju, Hong:2004sa}.

\section{The AdS/CFT Duality}

The essential principle
underlying the AdS/CFT approach to conformal gauge theories is the
isomorphism of the group of Poincare' and conformal transformations
$SO(4,2)$ to the group of isometries of Anti-de Sitter space.  The
AdS metric is
\begin{equation} \label{eq:AdSz}
ds^2 = \frac{R^2}{z^2}(\eta_{\mu \nu} dx^\mu
 dx^\nu - dz^2),
 \end{equation}
which is invariant under scale changes of the
coordinate in the fifth dimension $z \to \lambda z$ and $ x_\mu \to
\lambda x_\mu$.  Thus one can match scale transformations of the
theory in $3+1$ physical space-time to scale transformations in the
fifth dimension $z$.
In the AdS/CFT duality, the amplitude $\Phi(z)$ represents the
extension of the hadron into the additional fifth dimension.  The
behavior of $\Phi(z) \to z^\Delta$ at $z \to 0$ matches the
twist-dimension of the hadron at short distances $x^2 \to 0$.

\subsection{The Hard-Wall Holographic Model}

As shown by Polchinski
and Strassler~\cite{Polchinski:2001tt}, one can simulate confinement
by imposing boundary conditions in the holographic variable at
$z = z_0 ={1/\Lambda_{\rm QCD}}$. Confinement can also be introduced
by modifying the AdS metric to mimic a confining potential.
The resulting models, although {\it ad hoc}, provide a
simple semi-classical approximation to QCD which incorporate both constituent counting
rule behavior at short distances and confinement at large distances.

This simple approach, which has been described as a ``bottom-up''
approach, has been successful in obtaining general properties of
scattering amplitudes of hadronic bound states at strong
coupling~\cite{Polchinski:2001tt, Brodsky:2003px}, the low-lying
hadron spectra~\cite{Boschi-Filho:2002vd, deTeramond:2004qd,
deTeramond:2005su,  Hong:2006ta} and hadron couplings and chiral
symmetry breaking~\cite{Erlich:2005qh, DaRold:2005zs, Hirn:2005nr,
Ghoroku:2005vt}. The gauge theory/gravity duality also provides a
convenient framework for the description of deep inelastic
scattering structure functions at small $x$~\cite{Polchinski:2002jw}
and a unified description of hard and soft pomeron
physics~\cite{Brower:2006ea}. Recent applications to describe chiral
symmetry breaking~\cite{Babington:2003vm} and other meson and baryon
properties, have also been carried out within the framework of a
top-bottom approach to AdS/CFT using higher dimensional
branes~\cite{Peeters:2007ab}.

\subsection{The Soft-Wall Holographic Model}

The predicted mass spectrum in the truncated space hard-wall (HW) model is
linear $M \propto L$ at high orbital
angular momentum $L$, in contrast to the quadratic dependence $M^2 \propto L$ in the usual
Regge parameterization.
It has been shown recently that by  choosing a specific profile for a non-constant dilaton, the usual Regge  dependence  can be obtained~\cite{Karch:2006pv}. The procedure allows one
to retain conformal AdS metrics (\ref{eq:AdSz}) and to introduce a smooth cutoff  which depends on the profile of a dilaton background field $\varphi$
\begin{equation}
S = \int \! d^4x \, dz  \,\sqrt{g} \,e^{- \varphi(z) } \mathcal{L},
\end{equation}
where $\varphi$ is a function of the holographic coordinate $z$
which vanishes in the ultraviolet limit $z \to 0$. The IR hard-wall
or truncated space holographic model, discussed in the previous
section, corresponds to a constant dilaton field in the confining
region $\varphi(z) = \varphi_0$,  $z \leq 1/\Lambda_{\rm QCD}$, and
to very large values elsewhere: $\varphi(z) \to \infty$ for $z >
1/\Lambda_{\rm QCD}$. The introduction of a soft cutoff avoids the
ambiguities in the choice of boundary conditions at the infrared
wall. A convenient choice~\cite{Karch:2006pv} for the background
field with usual Regge behavior is $\varphi(z) = \kappa^2 z^2$. The
resulting wave equations are equivalent to the radial equation of a
two-dimensional oscillator previously found in the context  of mode
propagation on $AdS_5 \times S^5$ in the light-cone formulation of
Type II supergravity~\cite{Metsaev:1999kb}.

\section{Light-Front Fock Representation}

The light-front expansion is constructed by quantizing QCD
at fixed light-cone time~\cite{Dirac:1949cp} $\tau = t + z/c$ and
forming the invariant light-front Hamiltonian: $ H^{QCD}_{LF} = P^+
P^- - {\vec P_\perp}^2$ where $P^\pm = P^0 \pm
P^z$~\cite{Brodsky:1997de}.
The momentum generators $P^+$ and $\vec
P_\perp$ are kinematical; {\em i.e.}, they are independent of the
interactions. The generator $P^- = i {d\over d\tau}$ generates
light-cone time translations, and the eigen-spectrum of the Lorentz
scalar $ H^{QCD}_{LF}$ gives the mass spectrum of the color-singlet
hadron states in QCD: $H_{LC} \ket{\psi_h} = {\cal M}^2_h \ket{\psi_h}$.
The state $\ket{\psi_h}$ is an expansion in multi-particle Fock eigenstates
$\{\ket{n} \}$ of the free light-front
Hamiltonian:
$\vert \psi_h \rangle = \sum_n \psi_{n/h} \vert \psi_h \rangle $; the projection
of the eigensolution on the free Fock basis gives the hadronic
light-front wavefunctions. The
coefficients of the  Fock expansion $\psi_{n/h}(x_i, \mbf{k}_{\perp i}, \lambda_i)$
are independent of the total momentum $P^+$ and $\mbf{P}_{\! \perp}$ of
the hadron and depend only on the relative partonic coordinates,
the longitudinal momentum fraction $x_i = k_i^+/P^+$,
the relative transverse momentum $\mbf{k}_{\perp i}$ and $\lambda_i$,
the constituent's spin along the $z$ direction.
Momentum conservation requires
$\sum_{i=1}^n x_i =1$ and $\sum_{i=1}^n \mbf{k}_{\perp i}=0$.

\subsection{Light-Front Wavefunctions in Impact Space}

The holographic mapping of hadronic LFWFs to AdS string modes is
most transparent when one uses the impact parameter space
representation~\cite{Soper:1976jc}. The total position coordinate of
a hadron or its transverse center of momentum $\mbf{R}_\perp$, is
defined in terms of the energy momentum tensor $T^{\mu \nu}$
\begin{equation}
\mbf{R}_\perp = \frac{1}{P^+} \int dx^-
\negthinspace \int d^2 \mbf{x}_\perp \,T^{++} \,
\mbf{x}_\perp.
\end{equation}
In terms of partonic transverse coordinates
\begin{equation}
x_i \mbf{r}_{\perp i} = x_i \mbf{R}_\perp + \mbf{b}_{\perp i},
\end{equation}
where  the $\mbf{r}_{\perp i}$ are the physical transverse position
coordinates and  $\mbf{b}_{\perp i}$ frame independent  internal
coordinates, conjugate to the relative coordinates $\mbf{k}_{\perp i}$.
Thus, $\sum_{i=1}^n \mbf{b}_{\perp i} = 0$ and
$\mbf{R}_\perp = \sum_{i=1}^n  x_i \mbf{r}_{\perp i}$.
The LFWF $\psi_n(x_j, \mbf{k}_{\perp j})$ can be expanded in terms of the $n-1$ independent
coordinates $\mbf{b}_{\perp j}$,  $j = 1,2,\dots,n-1$
\begin{equation} \label{eq:LFWFb}
\psi_n(x_j, \mbf{k}_{\perp j}) =  (4 \pi)^\frac{(n-1)}{2}
\prod_{j=1}^{n-1}\int d^2 \mbf{b}_{\perp j}
\exp\!{\Big(i \sum_{j=1}^{n-1} \mbf{b}_{\perp j} \cdot \mbf{k}_{\perp j}\Big)}
\widetilde{\psi}_n(x_j, \mathbf{b}_{\perp j}).
\end{equation}
The normalization is defined by
\begin{equation}
\sum_n  \prod_{j=1}^{n-1} \int d x_j d^2 \mbf{b}_{\perp j}
\left\vert\widetilde \psi_n(x_j, \mbf{b}_{\perp j})\right\vert^2 = 1.
\end{equation}

One of the important advantages of the light-front formalism is that current
matrix elements can be represented without approximation as overlaps
of light-front wavefunctions. In the case of the elastic space-like
form factors, the matrix element of the $J^+$ current only couples
Fock states with the same number of constituents.  If
the charged parton $n$ is the active constituent struck by the
current, and the quarks $i = 1,2, \dots ,n-1$ are spectators, then
the Drell-Yan West formula~\cite{Drell:1969km,West:1970av,Brodsky:1980zm} in impact space is
\begin{equation} \label{eq:FFb}
F(q^2) =  \sum_n  \prod_{j=1}^{n-1}\int d x_j d^2 \mbf{b}_{\perp j}
\exp\!{\Bigl(i \mbf{q}_\perp \cdot \sum_{j=1}^{n-1} x_j \mbf{b}_{\perp j}\Bigr)}
\left\vert \widetilde \psi_n(x_j, \mbf{b}_{\perp j})\right\vert^2,
\end{equation}
corresponding to a change of transverse momenta $x_j \mbf{q}_\perp$
for each of the $n-1$ spectators.  This is a convenient form for
comparison with AdS results, since the form factor is expressed in
terms of the product of light-front wave functions with identical
variables.

\section{Light-Front Mapping of String Amplitudes}

We can now establish an explicit connection between the AdS/CFT and
the LF formulae. To make more transparent the holographic connection
between AdS$_5$ and the conformal quantum field theory defined at
its asymptotic $z\to 0$ boundary, it is convenient to write  the AdS
metric (\ref{eq:AdSz}) in terms of light front coordinates $x^\pm =
x^0 \pm x^3$
\begin{equation} \label{eq:AdSLF}
ds^2 = \frac{R^2}{z^2} \left( dx^+ dx^- - d \mbf{x}_\perp^2 - dz^2
\right).
\end{equation}
It is also useful to express (\ref{eq:FFb}) in terms of an effective
single particle transverse distribution $\widetilde \rho$
~\cite{Brodsky:2006uqa}
\begin{equation} \label{eq:FFzeta}
F(q^2) = 2 \pi \int_0^1 dx \frac{(1-x)}{x}  \int \zeta d \zeta\,
J_0\negthinspace\left(\zeta q \sqrt{\frac{1-x}{x}}\right) \tilde
\rho(x,\zeta),
\end{equation}
where we have introduced the variable
\begin{equation}
\zeta = \sqrt{\frac{x}{1-x}} ~\Big\vert \sum_{j=1}^{n-1} x_j
\mathbf{b}_{\perp j}\Big\vert,
\end{equation}
representing the $x$-weighted transverse impact coordinate of the
spectator system. On the other hand, the expression for the form
factor in AdS space is represented as the overlap in the fifth
dimension coordinate $z$ of the normalizable modes dual to the
incoming and outgoing hadrons, $\Phi_P$ and $\Phi_{P'}$, with the
non-normalizable mode, $J(Q,z) = z Q K_1(z Q)$, dual to the external
source~\cite{Polchinski:2002jw}
\begin{equation}
F(Q^2) = R^{3} \int \frac{dz}{z^{3}} \,
  \Phi_{P'}(z) J(Q,z) \Phi_P(z).
\label{eq:FFAdS}
\end{equation}
If we compare (\ref{eq:FFzeta}) in impact space with the expression
for the form factor in AdS space (\ref{eq:FFAdS}) for arbitrary
values of $Q$ using the identity
\begin{equation} \label{eq:intJ}
\int_0^1 dx \, J_0\negthinspace\left(\zeta Q
\sqrt{\frac{1-x}{x}}\right) = \zeta Q K_1(\zeta Q),
\end{equation}
then we can identify the spectator density function appearing in the
light-front formalism with the corresponding AdS density
\begin{equation} \label{eq:hc}
\tilde \rho(x,\zeta)
 =  \frac{R^3}{2 \pi} \frac{x}{1-x}
\frac{\left\vert \Phi(\zeta)\right\vert^2}{\zeta^4}.
\end{equation}
Equation (\ref{eq:hc}) gives a precise relation between  string
modes $\Phi(\zeta)$ in AdS$_5$ and the QCD transverse charge density
$\tilde\rho(x,\zeta)$. The variable $\zeta$ represents a measure of
the transverse separation between point-like constituents, and it is
also the holographic variable $z$ characterizing the string scale in
AdS. Consequently the AdS string mode $\Phi(z)$ can be regarded as
the probability amplitude to find $n$ partons at transverse impact
separation $\zeta = z$. Furthermore, its eigenmodes determine the
hadronic spectrum~\cite{Brodsky:2006uqa}.

In the case of a two-parton constituent bound state, the
correspondence between the string amplitude $\Phi(z)$ and the
light-front wave function $\widetilde\psi(x,\mathbf{b})$ is
expressed in closed form~\cite{Brodsky:2006uqa}
\begin{equation}  \label{eq:Phipsi}
\left\vert\widetilde\psi(x,\zeta)\right\vert^2 = \frac{R^3}{2 \pi}
~x(1-x)~ \frac{\left\vert \Phi(\zeta)\right\vert^2}{\zeta^4},
\end{equation}
where $\zeta^2 = x(1-x) \mathbf{b}_\perp^2$. Here $\mbf{b}_\perp$ is
the impact separation and Fourier conjugate to $\mbf{k}_\perp$.

In presence of the dilaton background $\varphi = \kappa^2 z^2$ the
form factor in AdS space has the additional term $e^{- \kappa^2
z^2}$  in the metric
\begin{equation} \label{eq:FFAdSSW}
F(Q^2) = R^3 \int \frac{dz}{z^3} \, e^{- \kappa^2 z^2} \Phi(z)
J_\kappa(Q,z) \Phi(z), 
\end{equation}
to be properly normalized to the charge operator at $Q = 0$. The
non-normalizable electromagnetic mode $J_\kappa(Q,z)$ in
(\ref{eq:FFAdSSW})
\begin{equation} \label{eq:Jkappa}
J_\kappa(Q,z) = \Gamma\!\left(1+ \frac{Q^2}{4 \kappa^2}\right)
U\!\left(\frac{Q^2}{4 \kappa^2}, 0 , \kappa^2 z^2\right),
\end{equation}
couples to the dilaton field as it propagates in AdS space and
$U(a,b,c)$ is the confluent hypergeometric function with the
integral representation
\begin{equation}
\Gamma(a)  U(a,b,z) =  \int_0^\infty \! e^{- z t} t^{a-1}
(1+t)^{b-a-1} dt.
\end{equation}

In the large $Q^2$ limit, $Q^2 \gg 4 \kappa^2$  we find that
$J_\kappa(Q,z) \to z Q K_1(z Q)$. Thus, for large transverse
momentum the current decouples from the dilaton background, and we
recover our previous results for the ultraviolet behavior of matrix
elements. When summed over all Fock states  the Drell-Yan-West
(DYW) formula gives an exact result. The formula describes the
coupling of the free electromagnetic current to the elementary
constituents in the interaction representation. In the presence of a
dilaton field in AdS space, the electromagnetic AdS mode is no
longer dual to a the free quark current, but  dual to a dressed
current, {\it i.e.}, a hadronic electromagnetic current including
virtual $\bar q q$ pairs and thus confined. We thus expect that the
modified mapping corresponds to the presence of higher Fock states
in the hadron.

\subsection{Holographic Light-Front Representation}

The mapping of $z$ from AdS space to $\zeta$ in the LF space  allows
the equations of motion in AdS space to be recast in the form of  a
light-front Hamiltonian equation~\cite{Brodsky:1997de}
\begin{equation}
H_{LF} \ket{\phi} = \mathcal{M}^2 \ket{\phi}, \label{eq:HLC}
\end{equation}
a remarkable result which  allows the discussion of the AdS/CFT
solutions in terms of light-front equations in physical 3+1
space-time. By substituting $\phi(\zeta) =
\left(\frac{\zeta}{R}\right)^{-3/2} \Phi(\zeta)$, in the AdS wave
equation describing the propagation of scalar modes in AdS space
\begin{equation} \label{eq:AdSPhiM}
\left[z^2 \partial_z^2 - (d-1) z\,\partial_z + z^2 \mathcal{M}^2 -
(\mu R)^2 \right] \Phi(z) = 0,
\end{equation}
we find an effective Schr\"odinger equation as a function of the
weighted impact variable $\zeta$
\begin{equation} \label{eq:Scheq}
\left[-\frac{d^2}{d \zeta^2} + V(\zeta) \right] \phi(\zeta) =
\mathcal{M}^2 \phi(\zeta),
\end{equation}
with the effective potential $V(\zeta) \to - (1-4 L^2)/4\zeta^2$ in
the conformal limit, where we identity $\zeta$ with the fifth
dimension $z$ of AdS space: $\zeta = z$. We have written above $(\mu
R)^2 = - 4 + L^2$. The solution to (\ref{eq:Scheq}) is $\phi(z) =
z^{-\frac{3}{2}} \Phi(z) = C z^\frac{1}{2} J_L(z \mathcal{M}).$ This
equation reproduces the AdS/CFT solutions for mesons with relative
orbital angular momentum $L$. The holographic hadronic light-front
wave functions $\phi(\zeta) = \langle \zeta \vert \phi \rangle$ are
normalized according to
\begin{equation}
\langle \phi \vert \phi \rangle = \int d\zeta \, \vert \langle
\zeta \vert \phi \rangle \vert^2 = 1,
\end{equation}
and represent the probability amplitude to find $n$-partons at
transverse impact separation $\zeta = z$. Its eigenmodes determine
the hadronic mass spectrum.

The effective wave equation
(\ref{eq:Scheq}) is a relativistic light-front equation defined at
$x^+ = 0$. The AdS metric $ds^2$ (\ref{eq:AdSLF}) is invariant if
$\mbf{x}_\perp^2 \to \lambda^2 \mbf{x}_\perp^2$ and $z \to \lambda
z$ at equal light-front time. The Casimir operator for the rotation
group $SO(2)$ in the transverse light-front plane is $L^2$. This
shows the natural holographic connection to the light front.

The lowest stable state $L = 0$ is determined by the
Breitenlohner-Freedman bound\cite{Breitenlohner:1982jf}.   Its
eigenvalues are set by the boundary conditions at $\phi(z =
1/\Lambda_{\rm QCD}) = 0$ and are given in terms of the roots of
Bessel functions: $\mathcal{M}_{L,k} = \beta_{L,k} \Lambda_{\rm
QCD}$. Normalized LFWFs $\widetilde\psi_{L,k}$ follow from
(\ref{eq:Phipsi})
\begin{equation}
\widetilde \psi_{L,k}(x, \zeta) =  B_{L,k} \sqrt{x(1-x)} J_L
\left(\zeta \beta_{L,k} \Lambda_{\rm QCD}\right) \theta\big(z \le
\Lambda^{-1}_{\rm QCD}\big),
\end{equation}
where $B_{L,k} = {\Lambda_{\rm QCD}}/ \sqrt{ \pi}
J_{1+L}(\beta_{L,k})$. The resulting wavefunctions depicted in Fig.
\ref{fig:MesonLFWF} display confinement at large interquark
separation and conformal symmetry at short distances, reproducing
dimensional counting rules for hard exclusive processes and the
scaling and conformal properties of the LFWFs at high relative
momenta in agreement  with perturbative QCD.
\begin{figure}[ht]
\centering
\includegraphics[angle=0,width=12.6cm]{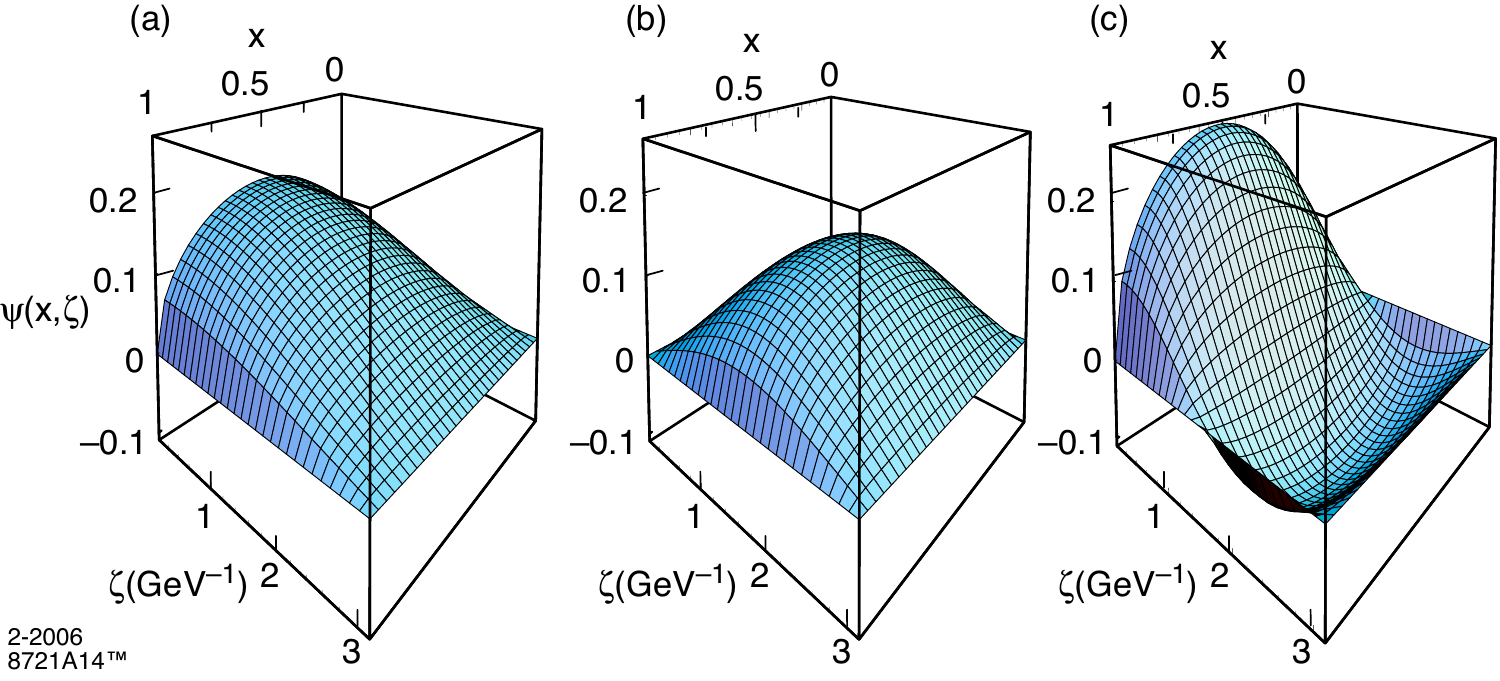}
\caption{AdS/QCD Predictions for the light-front wavefunctions of a
meson in the hard-wall model: (a) $n = 0, L = 0$, (b) $n = 0, L =
1$, (c)  $n = 1, L = 0$.} \label{fig:MesonLFWF}
\end{figure}

Since they are complete and orthonormal, these AdS/CFT model
wavefunctions can be used as an initial ansatz for a variational
treatment or as a basis for the diagonalization of the light-front
QCD Hamiltonian.  We are now in fact investigating this possibility
with J. Vary and A. Harinandrath. Alternatively, one can introduce
confinement by adding a two-dimensional harmonic oscillator potential $\kappa^4 \zeta^2$
in the LF transverse plane
to the conformal kernel in  Eq. (\ref{eq:Scheq}). One can also
introduce nonzero quark masses for the meson. The procedure is
straightforward in the $\mathbf{k}_\perp$ representation by using
the substitution $\frac{\mathbf{k}_\perp^2}{x(1-x)} \to
\frac{\mathbf{k}_\perp^2 +  m^2_1}{x} + \frac{ \mathbf{k}_\perp^2 +
m^2_2}{1-x}$.

\section{Integrability of AdS/CFT Equations}

The integrability methods of~cite{Infeld:1941} find a remarkable application in the AdS/CFT
correspondence.   Integrability follows
if the equations describing a physical model can be factorized in
terms of linear operators. These ladder operators
generate all the eigenfunctions once the lowest mass eigenfunction is known.
In holographic QCD, the conformally invariant 3 + 1 light-front differential equations can be expressed in terms of ladder operators and their solutions can then be expressed in terms of analytical functions.
In the conformal limit the ladder algebra for bosonic ($B$) or fermionic ($F$) modes is given in terms of
the operator ($\Gamma^B =1, ~~ \Gamma^F = \gamma_5$)
\begin{equation} \label{eq:Pi}
\Pi_\nu^{B,F}(\zeta) = -i \left( \frac{d}{d \zeta} - \frac{\nu + \half}{\zeta} \, \Gamma^{B,F} \!\right) ,
\end{equation}
and its adjoint
\begin{equation}
\Pi^{B,F}_\nu(\zeta)^\dagger = -i \left(\frac{d}{d \zeta} + \frac{\nu + \half}{\zeta}
\, \Gamma^{B,F} \!\right) ,
\end{equation}
with commutation relations
\begin{equation}
\left[\Pi_\nu^{B,F}(\zeta),\Pi_\nu^{B,F}(\zeta)^\dagger\right]
=  \frac{2 \nu+1}{\zeta^2} \, \Gamma^{B,F} .
\end{equation}
For $\nu \ge 0$ the Hamiltonian is written as a bilinear form
$H^{B,F}_{LC} = {\Pi_\nu^{B,F}}^\dagger \Pi_\nu^{B,F}$. In the fermionic case the eigenmodes
also satisfy a first order LF Dirac equation.
For bosonic modes, the lowest stable
state $\nu =0$ corresponds to the Breitenlohner-Freedman bound. Higher orbital states
are constructed from the $L$-th application of the raising operator $a^\dagger = - i \Pi^B$ on the
ground state.

\section{Hadronic Spectra in AdS/QCD}

The holographic model
based on truncated AdS space can be used to obtain the hadronic
spectrum of light quark $q \bar q, qqq$ and $gg$ bound states.  Specific
hadrons are identified by the correspondence of the amplitude in the fifth dimension with
the twist dimension of the interpolating operator for the hadron's valence
Fock state, including its orbital angular momentum excitations.   Bosonic modes with conformal
dimension $2+L$ are dual to the interpolating operator $\mathcal{O}_{\tau + L}$ with $\tau = 2$.
For fermionic modes $\tau = 3$.
For example,
the set of three-quark baryons with spin 1/2 and higher  is described
by the light-front Dirac equation
\begin{equation}
\left(\alpha \, \Pi^F \! (\zeta) - \mathcal{M} \right)  \psi(\zeta) = 0,
\end{equation}
where
$
 i \alpha =
  \begin{pmatrix}
  0& I\\
- I& 0
  \end{pmatrix}
$ in the Weyl representation.
The solution is\begin{equation} \label{eq:DiracLF}
\psi(\zeta) =  C \sqrt{\zeta}
\left[J_{L+1}\left(\zeta \mathcal{M} \right) \, u_+
+ J_{L+2}\left(z \mathcal{M}\right) \, u_- \right],
\end{equation}
with $\gamma_5 u_\pm = u_\pm$.
A discrete  four-dimensional spectrum follows when we impose the boundary condition
$\psi_\pm(\zeta=1/\Lambda_{\rm QCD}) = 0$:
$\mathcal{M}_{\alpha, k}^+ = \beta_{\alpha,k} \Lambda_{\rm QCD}, ~~
\mathcal{M}_{\alpha, k}^- = \beta_{\alpha + 1,k} \Lambda_{\rm QCD}$,
with a scale-independent mass ratio\cite{deTeramond:2005su}.

Figure \ref{fig:BaryonSpec}(a) shows the predicted orbital spectrum of the
nucleon states and Fig. \ref{fig:BaryonSpec}(b) the $\Delta$ orbital
resonances. The spin-3/2 trajectories are determined from the corresponding Rarita-Schwinger  equation.  The solution of the spin-3/2 for polarization along Minkowski coordinates, $\psi_\mu$, is similar to the spin-1/2 solution.
The data for the baryon spectra are from~\cite{Eidelman:2004wy}.
The internal parity of states is determined from the SU(6)
spin-flavor symmetry.
\begin{figure}[ht]
\centering
\includegraphics[angle=0,width=12.6cm]{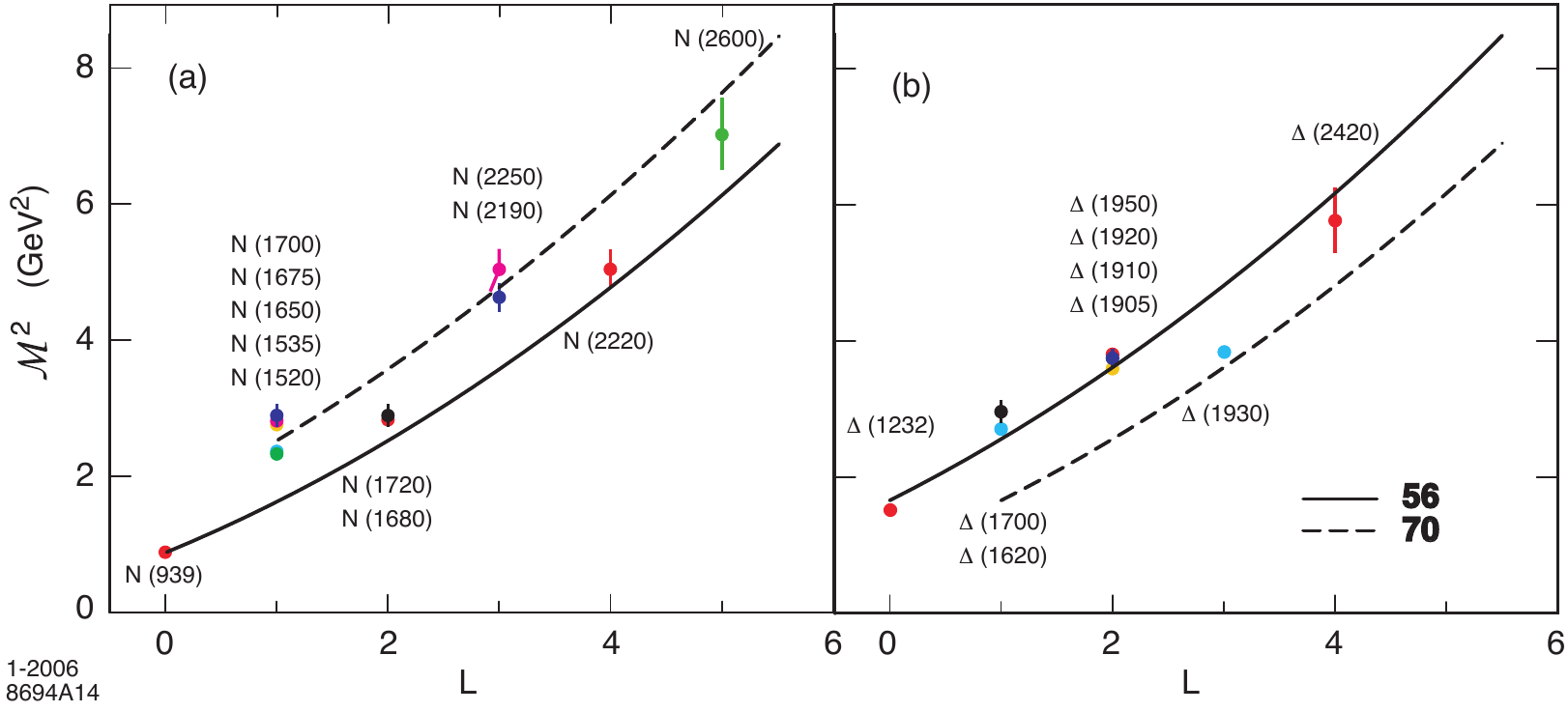}
\caption{Predictions for the light baryon orbital spectrum for
$\Lambda_{QCD}$ = 0.25 GeV. The  $\bf 56$ trajectory corresponds to
$L$ even  $P=+$ states, and the $\bf 70$ to $L$ odd  $P=-$ states.}
\label{fig:BaryonSpec}
\end{figure}
Since only
one parameter, the QCD mass scale $\Lambda_{\rm QCD}$, is introduced, the
agreement with the pattern of physical states is remarkable. In
particular, the ratio of $\Delta$ to nucleon trajectories is
determined by the ratio of zeros of Bessel functions.  The
predicted mass spectrum in the truncated space model is linear $M \propto L$ at high orbital
angular momentum, in contrast to the quadratic dependence $M^2
\propto L$ in the usual Regge parameterization. One can obtain $M^2 \propto (L+n)$ dependence
in the holographic model by the introduction of a harmonic potential $\kappa^2 z^2$
in the AdS wave equations~\cite{Karch:2006pv}.  This result can also be obtained by extending the conformal
algebra written above.
An account of the extended algebraic holographic model and a possible supersymmetric
connection between the bosonic and fermionic operators used in the holographic
construction will be described elsewhere.

\section{Analytic Results for Hadronic Observables in AdS/QCD}

\subsection{Hadronic Form Factors of Arbitrary Twist}

A string mode $\phi_\tau$ which couples to a local hadronic interpolating operator of twist $\tau$
defined at the asymptotic boundary of AdS space has scaling dimension $\tau$,
$\Phi_\tau \sim z^\tau$, as $z \to 0$.  In the soft-wall model a normalized string mode $\Phi_\tau$
is given by
\begin{equation}
\Phi_\tau(z) = \frac{1}{R^{3/2}} \sqrt{\frac{2}{\Gamma(\tau \! - \! 1)} } \, \kappa^{\tau -1} z ^{\tau}.
\end{equation}
for $n = 0$.  It is convenient to
perform the change of variable $ t = \frac{x}{1-x}$  in the integral representation for the
bulk-to-boundary propagator (\ref{eq:Jkappa}). Upon integration by parts the resulting expression,
there follows
\begin{equation} \label{eq:IntRepJx}
J_\kappa(Q,z) = \kappa^2 z^2 \int_0^1 \! \frac{dx}{(1-x)^2} \, x^{\frac{Q^2}{4 \kappa^2}}
e^{-\kappa^2 z^2 x/(1-x)},
\end{equation}
the result found by Grigoryan and Radyuskin in Ref.~\cite{Grigoryan:2007my}.
We can then  find an analytical expression for the form factor of a hadronic state of arbitrary twist $\tau$
by substituting the integral representation  (\ref{eq:IntRepJx}) for
$J_\kappa(q,z)$ in  (\ref{eq:FFAdSSW}). We find the result
\begin{equation} \label{eq:FFSWtau}
F(Q^2) = \Gamma(\tau)
\frac{\Gamma\left(1\! + \! \frac{Q^2}{4 \kappa^2}\right)}{\Gamma\left(\tau \! + \! \frac{Q^2}{4 \kappa^2}\right)}.
\end{equation}

In the absence of anomalous dimensions, the twist is an integer, $\tau = N$,   and we can simplify
(\ref{eq:FFSWtau}) by using the recurrence formula
\begin{equation}
\Gamma(N+z) = (N - 1 + z) (N - 2 + z) \dots (1 + z) \Gamma(1+z).
\end{equation}
We find
\begin{eqnarray} \label{eq:monopole}
 F(Q^2) \! &=& \! \frac{1}{1 + \frac{Q^2}{4 \kappa^2}},  ~~~N = 2,  ~~~\\ \label{eq:dipole}
F(Q^2) \! &=& \!  \frac{2}{\left(1  + \! \frac{Q^2}{4 \kappa^2}\right) \!
 \left(2 + \! \frac{Q^2}{4 \kappa^2}\right)},
~~~ N =  3, \\
  &   &  ~~~~~~~~   \cdots \nonumber \\  \label{eq:nminusonepole}
 F(Q^2) \!&=& \! \frac{(N-1)!}{\left(1 + \!\frac{Q^2}{4 \kappa^2} \right) \!
 \left(2 + \! \frac{Q^2}{4 \kappa^2}  \right)  \! \cdots \!
       \left(N \! - \! 1 \! + \! \frac{Q^2}{4 \kappa^2}  \right)},  \, N, ~~~~~~~
\end{eqnarray}
which is expressed as a product of $N-1$  poles, corresponding to the first $N-1$ states along the vector meson radial trajectory.
For large $Q^2$ it follows that
\begin{equation}
F(Q^2) \to (N-1)! \left[\frac{4 \kappa^2}{Q^2}\right]^{(N-1)},
\end{equation}
and we recover the conformal power-law counting rules for hard scattering~\cite{Brodsky:1974vy,Matveev:1973ra,Ferrara:1972qz}.

\subsection{Meson Form Factors}
As we have discussed,
hadron form factors can be predicted in AdS from the overlap of the normalizable modes dual to
the incoming and outgoing hadrons with the bulk-to-boundary propagator, $J(Q,z)$, dual to the external source, or equivalently by using the DYW formula in physical space-time.
For example, we can compute the pion form factor from the AdS expressions  (\ref{eq:FFAdS}) and (\ref{eq:FFAdSSW})
for the hadronic string modes $\Phi_\pi$ in the hard-wall (HW)
\begin{equation} \label{eq:PhipiHW}
\Phi_\pi^{HW}(z) = \frac{\sqrt{2} \Lambda_{QCD}}{R^{3/2} J_1(\beta_{0,1}) }
z^2 J_0\left(z \beta_{0,1} \Lambda_{QCD} \right) ,
\end{equation}
and soft-wall (SW) model
\begin{equation} \label{eq:PhipiSW}
\Phi_\pi^{SW}(z) = \frac{\sqrt{2} \kappa}{R^{3/2}}\, z^2 ,
\end{equation}
respectively.
Since the pion mode couples to a twist-two boundary
interpolating operator which creates a  two-component hadronic bound state, the
form factor is given in the SW model by the simple monopole form (\ref{eq:monopole}) corresponding
to $n=2$
\begin{equation} \label{eq:pionFF}
 F_\pi(Q^2) = \frac{4 \kappa^2}{4 \kappa^2 + Q^2}.
 \end{equation}
The hadronic scale is evaluated
 by fitting the space-like data for the form factor as shown in Figure \ref{fig:PionQ2FFSL}, where we plot
 the product $Q^2 F_\pi(Q^2)$ for the soft and hard-wall holographic models.  Both models would seem to describe the overall behavior of the space-like data; however, when  the low energy data is examined in detail, the SW model gives a noticeable better description as shown in Figure \ref{fig:PionFFSL}.
\begin{figure}[h]
\centering
\includegraphics[angle=0,width=8.6cm]{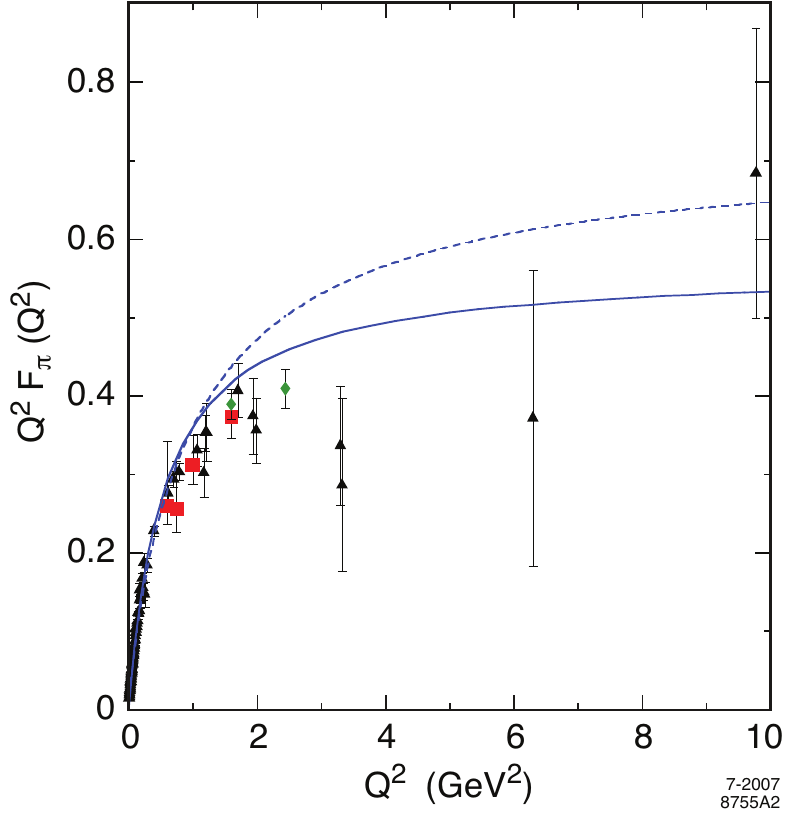}
\caption{Space-like scaling behavior for $Q^2 F_\pi(Q^2)$ as a function of $Q^2 = -q^2$. The continuous line is the prediction of the soft-wall model for  $\kappa = 0.375$ GeV. The dashed line is the prediction of the hard-model for $\Lambda_{QCD} = 0.22$ GeV. The black triangles is the data compilation  from Baldini  {\it et al.}~\cite{Baldini:1998qn},  the red boxes  are JLAB 1 data~\cite{Tadevosyan:2007yd} and cobalt green diamonds are JLAB 2  data~\cite{Horn:2006tm}. }
\label{fig:PionQ2FFSL}
\end{figure}
\begin{figure}[ h]
\centering
\includegraphics[angle=0,width=9.6cm]{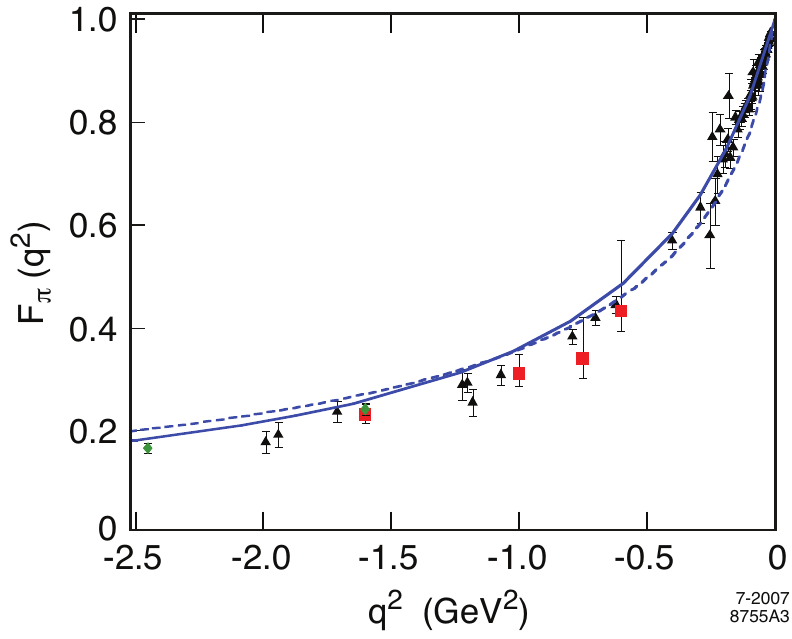}
\caption{Space-like behavior of the pion form factor $F_\pi(q^2)$ as a function of $q^2$ for $\kappa = 0.375$ GeV  and  $\Lambda_{QCD} = 0.22$ GeV. Continuous line: soft-wall model, dashed line: hard-wall model. Black (triangle) is the data compilation  from Baldini  {\it et al.}~\cite{Baldini:1998qn},  red (box)  is JLAB 1~\cite{Tadevosyan:2007yd} and cobalt green diamonds are JLAB 2~\cite{Horn:2006tm}.}
\label{fig:PionFFSL}
\end{figure}
When the results for the pion form factor are analytically continued
to the time-like region, $q^2 \to -q^2$ we obtain the results shown
in Figure \ref{fig:PionFFLog} for $\log\left(\vert
F_\pi(q^2)\vert\right)$. The monopole form of the SW model exhibits
a pole at the $\rho$ mass and reproduces  well the $\rho$ peak  with
$M_\rho = 4 \kappa^2 = 750$ MeV. In the strongly coupled
semiclassical gauge/gravity limit hadrons have zero widths  and are
stable. The form factor accounts for the scaling behavior in the
space-like region, but does not give rise to the additional
structure found in the time-like region, since the $\rho$ pole
saturates 100\% the monopole form (\ref{eq:pionFF}).
\begin{figure}[h ]
\centering
\includegraphics[angle=0,width=10.4cm]{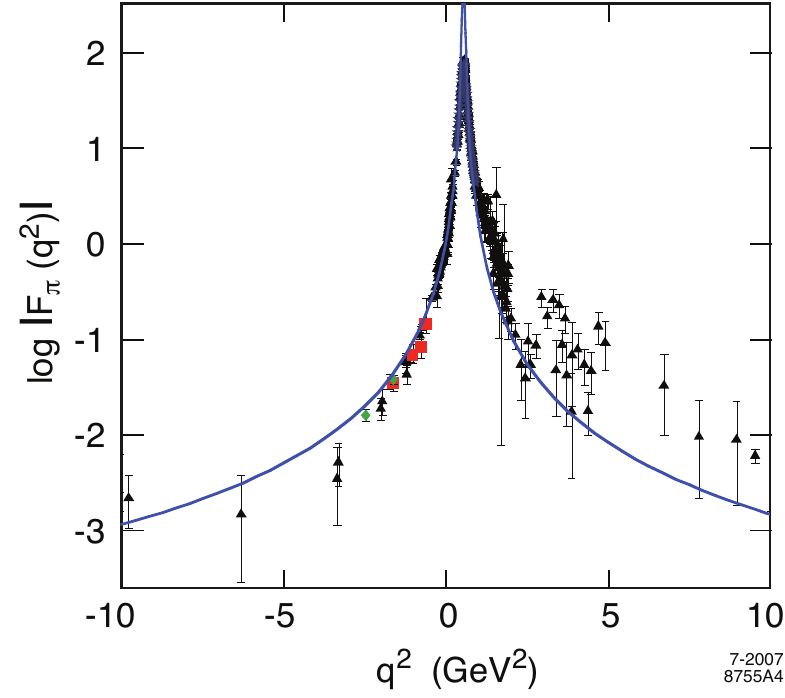}
\caption{Space and time-like behavior of the pion form factor $\log\left(\vert F_\pi(q^2)\vert\right)$ as a function of $q^2$ for $\kappa = 0.375$ GeV in the  soft-wall model. Black (triangle) is the data compilation  from Baldini  {\it et al.}~\cite{Baldini:1998qn},  red (box) is JLAB 1~\cite{Tadevosyan:2007yd} and cobalt green diamonds are JLAB 2~\cite{Horn:2006tm}.}
\label{fig:PionFFLog}
\end{figure}

\subsection{Contributions to Meson Form Factors and Structure
Functions at Large Momentum Transfer in AdS/QCD}

The  form factor of a hadron  at large $Q^2$ arises from the small
$z$ kinematic domain in  AdS space. According to the AdS/CFT
duality, this corresponds to small distances $x^\mu x_\mu \sim
{1/Q^2}$ in physical space-time, the domain where the current matrix
elements are controlled by the conformal twist-dimension, $\Delta$,
of the hadron's interpolating operator.  In the case of the front
form, where $x^+=0$,  this corresponds to small transverse
separation $x^\mu x_\mu =  -\mbf{x}^2_\perp.$

As we have shown~\cite{Brodsky:2006uqa}, one can use holography to
map the functional from of the string modes  $\Phi(z)$ in  AdS space
to the light front wavefunctions in physical  space time by
identifying $z$ with the  transverse variable $ \zeta =\sqrt{x\over
1-x} |\vec \eta_\perp | .$  Here  $\vec \eta_\perp =
\sum^{n-1}_{i=1}  x_i  \mbf{b}_{\perp i}$ is the weighted impact
separation,  summed over the impact separation of the spectator
constituents.  The leading large-$Q^2$  behavior of form factors in
AdS/QCD arises from small $\zeta \sim 1/ Q$,  corresponding to small
transverse separation.

For the case of a meson with two constituents the form factor can be
written in terms of an effective light-front transverse density in
impact space
\begin{equation} \label{eq:bxFF}
F(q^2) = \int^1_0 \! dx  \int  \!d b^2 \, \widetilde \rho(x, b, Q),
\end{equation}
with $\widetilde \rho(x, b, Q) = \pi J_0\left(b \,Q (1-x)\right) \vert \widetilde \psi(x, b) \vert^2$ and
$b = \vert \mbf{b}_\perp \vert$.   The kinematics are  illustrated  in figure 6 for the case of a meson with two constituents in the soft-wall model
\begin{equation}
\widetilde\psi_{q \bar q/\pi}(x, \mbf{b}_\perp)
= \frac{\kappa}{\sqrt{\pi}} \sqrt{x(1-x)}~e^{-\half \kappa^2 x(1-x) \mbf{b}_\perp^2},
\end{equation}
where the Gaussian form of the LFWF at large $Q^2$ favors
short-distance configurations with small $\zeta^2  = b^2_\perp
x(1-x) \sim 1/ Q^2$. Since we are mainly interested in studying the
contribution from different regions to  the form factor at large
$Q^2$, we have replaced the modified bulk-to-boundary propagator
$J_\kappa(Q, z)$ (\ref{eq:Jkappa}) by its large $Q^2$ form $z Q
K_1(z Q)$. One sees a shift of the integrand $\tilde  \rho(x, b, Q)$
toward small $\vert \mbf{b}_\perp \vert$ and small $1-x$ at high
$Q^2.$  A similar behavior is observed for the LFWF obtained from
the hard wall model.
\begin{figure}[h]  \label{fig:bxFF}
\centering
\includegraphics[angle=0,width=10.6cm]{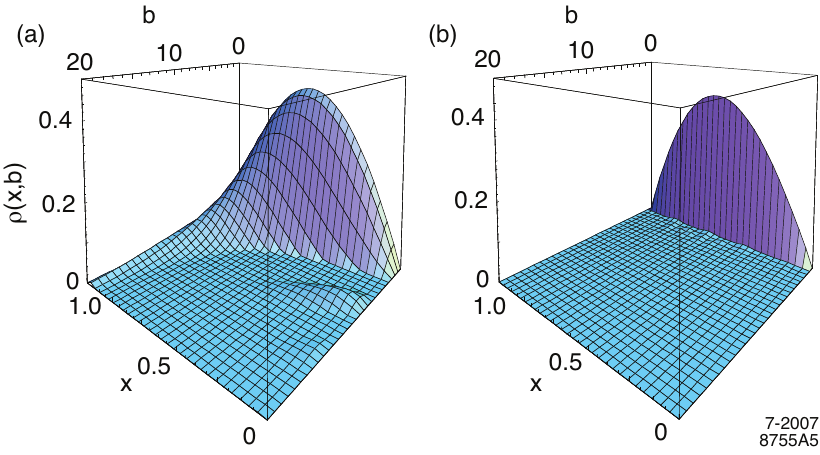}
\caption{Effective partonic density $ \rho(x,b,Q)$ in terms of the
 longitudinal momentum fraction $x$, the transverse relative impact
variable $b = \vert \mbf{b}_\perp\vert$ and momentum transfer $Q$ for the soft
wall model. As $Q$ increases the distribution becomes increasingly important near
$x=1$ and $\mbf{b}_\perp =0$. This is illustrated in (a) for $Q$ = 1 GeV/c.
At very large $Q$ (figure (b)), the distribution is peaked
towards $\mbf{b}_\perp = 0$.
The value of $\kappa$ is 0.375 GeV.}
\end{figure}

\subsection{The Meson Structure Function}

The pion structure function $q_\pi(x,Q^2)$ is computed by integrating the square of the
pion light-front wave function up to the scale $Q^2$
\begin{equation}
q_\pi(x,Q^2) = \int^{Q^2} \frac{d^2 \mbf{k}_\perp}{16 \pi^3} ~
\left\vert \psi_{\bar q q /\pi}(x, \mbf{k}_\perp)\right\vert^2,
\end{equation}
where the LFWF in $\mbf{k}_\perp$ space
\begin{equation}  \label{eq:LFWFkSW}
\psi_{\bar q q/\pi}(x, \mbf{k}_\perp) = \frac{4 \pi}{\kappa \sqrt{x(1-x)}}
~e^{- \frac{\mbf{k}_\perp^2}{2 \kappa^2 x(1-x)}} .
\end{equation}
is the Fourier transform (\ref{eq:LFWFb}) for a two-parton bound state.
We find
\begin{equation}
q_\pi(x,Q^2) = 1 - e^{-\frac{Q^2}{\kappa^2 x(1-x)}},
\end{equation}
In the large $Q^2$ limit $q_\pi(x,Q^2 \to \infty) \equiv q_\pi(x) = 1$, which is the behavior of a strongly coupled theory found  in QCD(1+1)~\cite{Pauli:1985ps}.  Identical results~\cite{Radyushkin:2006iz} are obtained for the pion in the hard wall model.

\subsection{The Pion Decay Constant}
The pion decay constant is given by the matrix element of the axial isospin
current $J^{\mu 5 a}$ between a physical pion and the vacuum state~\cite{Peskin:1995ev}
$\bigl\langle 0 \bigl\vert J_W^+(0) \bigr\vert \pi^-(P^+, \vec P_\perp) \bigr\rangle$,
where $J_W^+$ is the flavor changing weak current.
Only the valence state
with $L_z = 0$, $S_z = 0$, contributes to the decay of the $\pi^\pm$.
Expanding the  hadronic initial state in the decay amplitude
into its Fock components we find
 \begin{equation}
f_\pi =  2 \sqrt{N_C} \int_0^1 dx \int \frac{d^2 \vec k_\perp}{16 \pi^3}
~\psi_{\bar q q/\pi}(x,k_\perp).
\end{equation}
This light-cone equation allows the exact computation of the pion decay
constant in terms of the valence pion light-front wave function~\cite{Lepage:1980fj}.

\subsection{The Meson Distribution Amplitude}
The meson distribution amplitude $\phi(x,Q)$ is defined as~\cite{Lepage:1979zb}
\begin{equation}
\phi(x,Q) = \int^{Q^2} \frac{d^2 \mbf{k}_\perp}{16 \pi^3}~\psi(x, \mbf{k}_\perp).
\end{equation}
It follows that
\begin{equation} \label{eq:phix}
\phi_\pi(x, Q\to\infty)
= \frac{4}{\sqrt{3}\pi}  f_\pi \sqrt{x(1-x)},
\end{equation}
with
\begin{equation}
f_\pi = \frac{1}{8} \sqrt{\frac{3}{2}} \, R^{3/2} \lim_{\zeta \to 0}
\frac{\Phi(\zeta)}{\zeta^2},
\end{equation}
since $\phi(x, Q \to \infty) \to
\widetilde \psi(x,\mbf{b}_\perp \to 0)/\sqrt{4 \pi}$ and
$\Phi_\pi \sim \zeta^2$ as $\zeta\to 0$.
The pion decay constant depends only on the behavior of the AdS
string mode near the asymptotic boundary, $\zeta = z = 0$ and the
mode normalization. For the truncated-space (TS) pion mode we find
$
f_\pi^{TS} = \frac{\sqrt{3}}{8 J_1(\beta_{0,k})} \, \Lambda_{\rm QCD}
= 83. 4 ~{\rm MeV},$
for $\Lambda_{QCD} = 0.22$ MeV. The corresponding result for the
transverse harmonic oscillator (HO) pion mode is
$
f_\pi^{HO} =  \frac{\sqrt{3}}{8} \, \kappa = 86.6 ~{\rm MeV},
$
for $\kappa = 0.375$ GeV.  The values of $\Lambda_{QCD}$ and $\kappa$ are
determined from the space-like form factor data as discussed above.
The experimental result for $f_\pi$ is extracted from the rate
of weak $\pi$ decay and has the value
$f_\pi = 92.4$ MeV~\cite{Eidelman:2004wy}.

It is interesting to note that the pion distribution amplitude
predicted by AdS/QCD (\ref{eq:phix})
has a quite different $x$-behavior than the
asymptotic distribution amplitude predicted from the PQCD
evolution~\cite{Lepage:1979zb} of the pion distribution amplitude
$\phi_\pi(x,Q \to \infty)= \sqrt 3  f_\pi x(1-x) $.  The broader
shape of the pion distribution increases the magnitude of the
leading twist perturbative QCD prediction for the pion form factor
by a factor of $16/9$ compared to the prediction based on the
asymptotic form, bringing the NLO PQCD prediction  close to the
empirical pion form factor~\cite{Choi:2006ha}.  The Fermilab E791
measurements~\cite{Aitala:2000hb} of the $x$ and $k_\perp$ fall-off of the diffractive dijet cross section $\pi A \to {\it Jet} ~  {\it Jet}  ~A $ also suggest a broad pion distribution amplitude for jet transverse momenta  $k^2_\perp  < {2 ~\rm GeV}^2.$

\subsection{The Nucleon Dirac Form Factors}
As a final example we compute the spin non-flip nucleon form factor in the soft wall
model. Consider the spin non-flip form factors
 \begin{eqnarray}
 F_+(Q^2) &\!=\!&  g_+\int \frac{d z}{z}  \, e^{- \kappa^2 z^2} J_\kappa(Q, z)
  \vert \Psi_+(z)\vert^2 ,\\
 F_-(Q^2) &\!=\!&  g_-\int  \frac{d z}{z} \, e^{- \kappa^2 z^2} J_\kappa(Q, z)
  \vert \Psi_-(z)\vert^2 ,
 \end{eqnarray}
 where the effective charges $g_+$ and $g_-$ are determined from the spin-flavor structure of
 the theory.
 We choose the struck quark to have $S^z=+1/2$. The two AdS solutions $\Psi_+$ and $\Psi_-$
correspond to nucleons with total angular momentum $J^z= +1/2$ and $-1/2$.
For the $SU(6)$ spin-flavor symmetry
\begin{eqnarray}
F_1^p(Q^2) &\!=\!&  \int  \frac {d z}{z}  \, e^{- \kappa^2 z^2} J_\kappa(Q, z)
  \vert \Psi_+(\zeta)\vert^2 ,\\
F_1^n(Q^2) &\!=\!& - \frac{1}{3}  \int  \frac{d z}{z}  \, e^{- \kappa^2 z^2} J_\kappa(Q, z)
 \left[\vert \Psi_+(z)\vert^2 - \vert\Psi_-(z)\vert^2\right],
 \end{eqnarray}
where $F_1^p(0) = 1$,~ $F_1^n(0) = 0$. The bulk-to-boundary
propagator $J_\kappa(Q,z)$ is the solution (\ref{eq:Jkappa}) of the
AdS wave equation for the external electromagnetic current,  and the
plus and minus components of the twist 3 nucleon mode in the soft
wall model are
\begin{eqnarray} \label{eq:PhipiSW}
\Psi_+(z) \!=\! \frac{\sqrt{2} \kappa^2}{R^{3/2}}\, z^3, \\
\Psi_-(z) \!=\! \frac{ \kappa^3}{R^{3/2}}\, z^4.
\end{eqnarray}
The results for $Q^4 F_1^p(Q^2)$ and $Q^4 F_1^n(Q^2)$ follow from the analytic form for the
form factors for any $\tau$ given in Section 7.1 and are shown in
Figure \ref{fig:nucleonFF}.
\begin{figure}[h]
\centering
\includegraphics[angle=0,width=8.8cm]{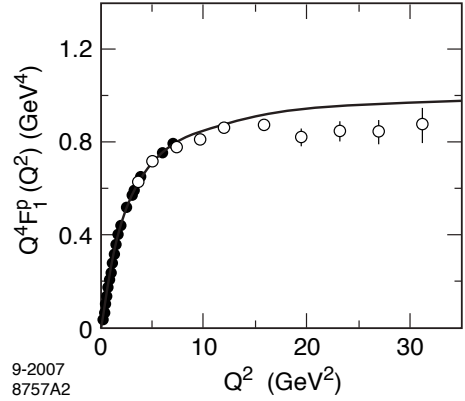}
\includegraphics[angle=0,width=8.8cm]{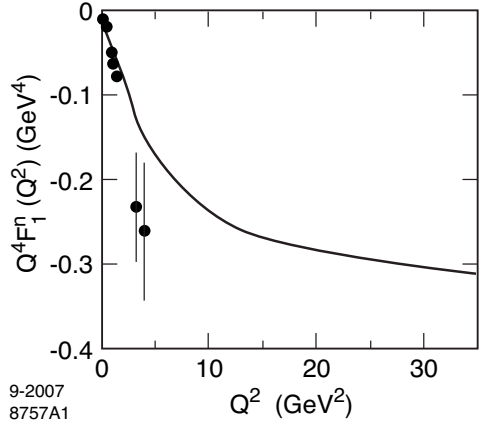}
\caption{Predictions for $Q^4 F_1^p(Q^2)$ and $Q^4 F_1^n(Q^2)$ in the
soft wall model for $\kappa =  0.424$ GeV.  The data compilation is
from Diehl~\cite{Diehl:2005wq}.} \label{fig:nucleonFF}
\end{figure}

\section{Overview of Hard Exclusive Processes}

The leading power fall-off of the hard scattering amplitude follows from the conformal scaling of
the underlying hard-scattering amplitude: $T_H \sim 1/Q^{n-4}$, where $n$ is
the total number of fields (quarks, leptons, or gauge fields)
participating in the hard
scattering~\cite{Brodsky:1974vy,Matveev:1973ra}. Thus the reaction
is dominated by subprocesses and Fock states involving the minimum
number of interacting fields.  In the case of $2 \to 2$ scattering
processes, this implies $ {d\sigma/ dt}(A B \to C D) ={F_{A B \to C
D}(t/s)/ s^{n-2}},$ where $n = N_A + N_B + N_C +N_D$ and $n_H$ is
the minimum number of constituents of $H$.   These dimensional counting rules are also a key feature of AdS/CFT~\cite{Polchinski:2001tt}.
The near-constancy of the
effective QCD coupling at small scales helps explain the general empirical
success of the dimensional counting rules for the near-conformal power
law fall-off of form factors and fixed-angle
scaling~\cite{Brodsky:1989pv}.  For example, measurements of pion photoproduction
are consistent with dimensional counting $s^{7}{ d\sigma/dt}(\gamma p \to \pi^+  n) \sim $ constant
at fixed CM angle for $s > 7 $ GeV.   The angular distributions seen in hard  large CM angle scattering reactions are consistent with quark interchange, a result predicted by the hard wall AdS/QCD model.
Reviews are given in refs.~\cite{Sivers:1975dg} and~\cite{White:1994tj}.
One sees the onset of perturbative QCD scaling behavior even for
exclusive nuclear amplitudes such as deuteron photodisintegration
(Here $n = 1+ 6 + 3 + 3 = 13 $) $s^{11}{ d\sigma/dt}(\gamma d \to p
n) \sim $ constant at fixed CM
angle~\cite{Holt:1990ze,Bochna:1998ca,Rossi:2004qm}. The measured
deuteron form factor~\cite{Rock:1991jy} also appears to follow the
leading-twist QCD predictions~\cite{Brodsky:1976rz} at large
momentum transfers in the few GeV region.  A measurement of ${
d\sigma/dt}(\gamma d \to \Delta^{++}\Delta^)$  in the scaling region
can establish the role of ``hidden-color" degrees of
freedom~\cite{Brodsky:1983vf} of the nuclear wavefunction in hard
deuteron reactions.

Recently the Hall A collaboration~\cite{Danagoulian:2007gs} at JLab
has reported a significant exception to the general empirical
success of dimensional counting in fixed CM angle Compton scattering
${d\sigma/dt}(\gamma p \to \gamma p) \sim {F(\theta_{CM})/s^8} $
instead of the predicted $1/s^6$ scaling.  The deviations from
fixed-angle conformal scaling may be due to corrections from
resonance contributions in the JLab energy range. It is interesting
that the hadron form factor $R_V(t)$~\cite{Diehl:1998kh}, which
multiplies the $\gamma q \to \gamma q$ amplitude is found by Hall-A
to scale as $1/t^2$, in agreement with the  PQCD and AdS/CFT
prediction. In addition the Belle measurement~\cite{Chen:2001sm} of
the timelike two-photon cross section ${d \sigma/dt}(\gamma \gamma
\to p \bar p)$ is consistent with $1/ s^6$ scaling.

Although large-angle proton-proton elastic scattering is well
described by dimensional scaling $s^{10}{ d\sigma/dt}(p p \to p p)
\sim $ constant at fixed CM angle,  extraordinarily  large spin-spin
correlations are observed~\cite{Court:1986dh}. The ratio of
scattering cross sections for spin-parallel and normal to the
scattering plane versus spin-antiparallel reaches $R_{NN} \simeq 4$
in large angle $ p p \to p p$ at $\sqrt s \simeq 5~$GeV; this is a
remarkable example of ``exclusive transversity".  Color transparency
is observed at lower energies but it fails~\cite{Mardor:1998zf} at
the same energy where $R_{NN}$ becomes large.  In fact, these
anomalies have a natural explanation~\cite{Brodsky:1987xw} as a
resonance effect related to the charm threshold in $pp$ scattering.
Alternative explanations of the large spin correlation are discussed
and reviewed in ref.~\cite{Dutta:2004fw}.
Resonance formation is a natural phenomenon when all constituents
are relatively at rest. For example, a resonance effect can occur
due to the intermediate state  $uud uud c \bar c$ at the charm
threshold $\sqrt s = 5$ GeV in $p p$ collisions. Since the $c $ and
$\bar c$ have opposite intrinsic parity, the resonance appears in
the $L= J= S =1$ partial wave for $ p p \to p p$ which is only
allowed for spin-parallel and normal scattering
$A_{NN}=1$~\cite{Brodsky:1987xw}. Resonance formation at the charm
threshold also explains the dramatic quenching of color transparency
seen in quasielastic $p n$ scattering by the EVA BNL
experiment~\cite{Mardor:1998zf} in the same kinematic region. The
reason why these effects are so dramatic in $p p \to p p$ scattering
is that the amplitude for the formation of an $uud uud c \bar c$
$s$-channel resonance in the intermediate state is of the same
magnitude as the fast-falling background $p p \to  p p$ PQCD
amplitude from quark interchange at large CM angles: $M(p p \to pp)
\sim {1/u^2 t^2} .$   We have predicted that the open charm cross
section in $p p$ scattering  will be of order of $1 ~\mu b$ at
threshold~\cite{Brodsky:1987xw}.

Color transparency~\cite{Brodsky:1988xz,Bertsch:1981py} is a key
property of color gauge theory, and it thus stands at the
foundations of QCD. Color transparency has been confirmed in
diffractive dijet production~\cite{Aitala:2000hc}, pion
photoproduction~\cite{Clasie:2007gq} and vector meson
electroproduction~\cite{Airapetian:2002eh}, but it is very important
to also systematically validate it in large angle hadron scattering
processes. One also expects similar novel QCD phenomena in
large-angle photoproduction $\gamma p \to \pi N$ near the charm
threshold, including the breakdown of color transparency and strong
spin-spin correlations. These effects  can be tested by measurements
at the new JLab 12 GeV facility,  which would confirm resonance
formation in a low partial wave in $\gamma p \to \pi N$ at $\sqrt s
\simeq 4$ GeV due to attractive forces in the $uu d \bar c c$
channel.

\section{Conclusions}

We have shown how the AdS/CFT correspondence between Anti-de Sitter
space and conformal gauge provides an analytically tractable
approximation to QCD in the regime where the QCD coupling is large
and constant. In particular, there is an exact correspondence
between the fifth-dimension coordinate $ z $ of AdS space and an
impact variable $\zeta$ which measures the invariant separation of
the quark constituents within the hadron in ordinary space-time.
This connection allows one to compute the analytic form of the
frame-independent light-front wavefunctions of mesons and baryons,
the fundamental entities which encode hadron properties and allow
the computation of exclusive scattering amplitudes.

The phenomenology of the AdS/QCD model is just beginning, but it can
be anticipated that it will have many applications to hadron
dynamics. For example, the model LFWFs obtained from AdS/QCD provide
a basis for understanding hadron structure functions and
fragmentation functions at the amplitude level; the same
wavefunctions also describe hadron formation from the coalescence of
co-moving quarks.  The spin correlations which underly single and
double spin correlations are also described by the AdS/QCD
eigensolutions.  The AdS/QCD hadronic wavefunctions provide
predictions for the generalized parton distributions of hadrons and
their weak decay amplitudes from first principles.  The amplitudes
relevant to diffractive reactions could also be computed. We also
anticipate that the extension of the AdS/QCD formalism to heavy
quarks will allow a great variety of heavy hadron phenomena to be
analyzed from first principles.

\vspace{10pt}

\noindent{\bf Acknowledgments}

\vspace{5pt}

This research was supported by the Department of Energy contract
DE--AC02--76SF00515. The research of GdT is supported in part by an
Aportes grant from Florida Ice \& Farm. Invited talk, presented at the 
Workshop on Exclusive Reactions at High Momentum Transfer
21-24 May 2007,
Newport News, Virginia.


\begin{thebibliography}{9}

\bibitem{Dirac:1949cp}
  P.~A.~M.~Dirac,
  ``Forms Of Relativistic Dynamics,''
  Rev.\ Mod.\ Phys.\  {\bf 21}, 392 (1949).

\bibitem{Brodsky:1998hn}
  S.~J.~Brodsky and D.~S.~Hwang,
  ``Exact light-cone wavefunction representation of matrix elements of
  electroweak currents,''
  Nucl.\ Phys.\  B {\bf 543}, 239 (1999)
  [arXiv:hep-ph/9806358].

\bibitem{Brodsky:2000xy}
  S.~J.~Brodsky, M.~Diehl and D.~S.~Hwang,
  ``Light-cone wavefunction representation of deeply virtual Compton
  scattering,''
  Nucl.\ Phys.\  B {\bf 596}, 99 (2001)
  [arXiv:hep-ph/0009254].

\bibitem{Lepage:1979zb}
  G.~P.~Lepage and S.~J.~Brodsky,
  ``Exclusive Processes In Quantum Chromodynamics: Evolution Equations For
  Hadronic Wave Functions And The Form-Factors Of Mesons,''
  Phys.\ Lett.\  B {\bf 87}, 359 (1979).

\bibitem{Efremov:1979qk}
  A.~V.~Efremov and A.~V.~Radyushkin,
  ``Factorization And Asymptotical Behavior Of Pion Form-Factor In QCD,''
  Phys.\ Lett.\  B {\bf 94}, 245 (1980).

\bibitem{Lepage:1980fj}
  G.~P.~Lepage and S.~J.~Brodsky,
  ``Exclusive Processes In Perturbative Quantum Chromodynamics,''
  Phys.\ Rev.\  D {\bf 22}, 2157 (1980).

  \bibitem{Brodsky:1979qm}
  S.~J.~Brodsky and G.~P.~Lepage,
  ``Exclusive Processes And The Exclusive Inclusive Connection In Quantum
  Chromodynamics,''
  SLAC-PUB-2294 (1979).

\bibitem{Maldacena:1997re}
  J.~M.~Maldacena,
  ``The large N limit of superconformal field theories and supergravity,''
  Adv.\ Theor.\ Math.\ Phys.\  {\bf 2}, 231 (1998)
  [Int.\ J.\ Theor.\ Phys.\  {\bf 38}, 1113 (1999)]
  [arXiv:hep-th/9711200];
  S.~S.~Gubser, I.~R.~Klebanov and A.~M.~Polyakov,
  ``Gauge theory correlators from non-critical string theory,''
  Phys.\ Lett.\ B {\bf 428}, 105 (1998)
  [arXiv:hep-th/9802109];
  E.~Witten,
  ``Anti-de Sitter space and holography,''
  Adv.\ Theor.\ Math.\ Phys.\  {\bf 2}, 253 (1998)
  [arXiv:hep-th/9802150].


\bibitem{vonSmekal:1997is}
  L.~von Smekal, R.~Alkofer and A.~Hauck,
  ``The infrared behavior of gluon and ghost propagators in Landau gauge
  QCD,''
  Phys.\ Rev.\ Lett.\  {\bf 79}, 3591 (1997)
  [arXiv:hep-ph/9705242].

\bibitem{Furui:2006py}
  S.~Furui and H.~Nakajima,
  ``Infrared features of unquenched finite temperature lattice Landau gauge
  QCD,''
  arXiv:hep-lat/0612009;
 S.~Furui and H.~Nakajima,
  ``Roles of the quark field in the infrared lattice Coulomb gauge and Landau
  gauge QCD,''
  arXiv:0708.1421 [hep-lat].

\bibitem{Brodsky:2002nb}
  S.~J.~Brodsky, S.~Menke, C.~Merino and J.~Rathsman,
  ``On the behavior of the effective QCD coupling alpha(tau)(s) at low
  scales,''
  Phys.\ Rev.\  D {\bf 67}, 055008 (2003)
  [arXiv:hep-ph/0212078].

\bibitem{Deur:2005cf}
  A.~Deur, V.~Burkert, J.~P.~Chen and W.~Korsch,
  ``Experimental determination of the effective strong coupling constant,''
  Phys.\ Lett.\  B {\bf 650}, 244 (2007)
  [arXiv:hep-ph/0509113].

\bibitem{Cornwall:1981zr}
  J.~M.~Cornwall,
  ``Dynamical Mass Generation In Continuum QCD,''
  Phys.\ Rev.\  D {\bf 26}, 1453 (1982).

   \bibitem{Brodsky:2006uqa}
  S.~J.~Brodsky and G.~F.~de Teramond,
  ``Hadronic spectra and light-front wavefunctions in holographic QCD,''
  Phys.\ Rev.\ Lett.\  {\bf 96}, 201601 (2006)
  [arXiv:hep-ph/0602252].

 \bibitem{deTeramond:2006xb}
  G.~F.~de Teramond,
  ``Mapping string states into partons: Form factors and the hadron spectrum in
  arXiv:hep-ph/0606143.

\bibitem{Grigoryan:2007vg}
  H.~R.~Grigoryan and A.~V.~Radyushkin,
  ``Form Factors and Wave Functions of Vector Mesons in Holographic QCD,''
  Phys.\ Lett.\  B {\bf 650}, 421 (2007)
  [arXiv:hep-ph/0703069];
  ``Pion Form Factor in Chiral Limit of Hard-Wall AdS/QCD Model,''
  arXiv:0709.0500 [hep-ph];
  H.~R.~Grigoryan,
  ``Dimension Six Correction to the Vector Sector of AdS/QCD Model,''
  arXiv:0709.0939 [hep-ph].

  \bibitem{Grigoryan:2007my}
  H.~R.~Grigoryan and A.~V.~Radyushkin,
  ``Structure of Vector Mesons in Holographic Model with Linear Confinement,''
  arXiv:0706.1543 [hep-ph].

\bibitem{Brodsky:2007pt}
  S.~J.~Brodsky and G.~F.~de Teramond,
  ``AdS/CFT and QCD,''
  arXiv:hep-th/0702205.

  \bibitem{Brodsky:2007hb}
  S.~J.~Brodsky and G.~F.~de Teramond,
  ``Light-Front Dynamics and AdS/QCD: The Pion Form Factor in the Space- and
  Time-Like Regions,''
  arXiv:0707.3859 [hep-ph].

  \bibitem{Huang:2007uu}
  T.~Huang and F.~Zuo,
  ``Couplings of the Rho Meson in a Holographic dual of QCD with Regge
  Trajectories,''
  arXiv:0708.0936 [hep-ph].

  \bibitem{Kwee:2007dd}
  H.~J.~Kwee and R.~F.~Lebed,
  ``Pion Form Factors in Holographic QCD,''
  arXiv:0708.4054 [hep-ph].

 \bibitem{Polchinski:2001ju}
  J.~Polchinski and L.~Susskind,
  ``String theory and the size of hadrons,''
  arXiv:hep-th/0112204.

\bibitem{Hong:2004sa}
  S.~Hong, S.~Yoon and M.~J.~Strassler,
  ``On the couplings of vector mesons in AdS/QCD,''
  JHEP {\bf 0604}, 003 (2006)
  [arXiv:hep-th/0409118].

  \bibitem{Polchinski:2001tt}
  J.~Polchinski and M.~J.~Strassler,
  ``Hard scattering and gauge/string duality,''
  Phys.\ Rev.\ Lett.\  {\bf 88}, 031601 (2002)
  [arXiv:hep-th/0109174].

\bibitem{Brodsky:2003px}
  S.~J.~Brodsky and G.~F.~de Teramond,
  ``Light-front hadron dynamics and AdS/CFT correspondence,''
  Phys.\ Lett.\  B {\bf 582}, 211 (2004)
  [arXiv:hep-th/0310227].

  \bibitem{Boschi-Filho:2002vd}
  H.~Boschi-Filho and N.~R.~F.~Braga,
  ``Gauge / string duality and scalar glueball mass ratios,''
  JHEP {\bf 0305}, 009 (2003)
  [arXiv:hep-th/0212207].

\bibitem{deTeramond:2004qd}
  G.~F.~de Teramond and S.~J.~Brodsky,
  ``Baryonic states in QCD from gauge / string duality at large N(c),''
  arXiv:hep-th/0409074.

\bibitem{deTeramond:2005su}
  G.~F.~de Teramond and S.~J.~Brodsky,
  ``The hadronic spectrum of a holographic dual of QCD,''
  Phys.\ Rev.\ Lett.\  {\bf 94}, 201601 (2005)
  [arXiv:hep-th/0501022].

  \bibitem{Hong:2006ta}
  D.~K.~Hong, T.~Inami and H.~U.~Yee,
  ``Baryons in AdS/QCD,''
  Phys.\ Lett.\  B {\bf 646}, 165 (2007)
  [arXiv:hep-ph/0609270].

\bibitem{Erlich:2005qh}
  J.~Erlich, E.~Katz, D.~T.~Son and M.~A.~Stephanov,
  ``QCD and a holographic model of hadrons,''
  Phys.\ Rev.\ Lett.\  {\bf 95}, 261602 (2005)
  [arXiv:hep-ph/0501128].

   \bibitem{DaRold:2005zs}
  L.~Da Rold and A.~Pomarol,
   ``Chiral symmetry breaking from five dimensional spaces,''
  Nucl.\ Phys.\ B {\bf 721}, 79 (2005)
  [arXiv:hep-ph/0501218];
  ``The scalar and pseudoscalar sector in a five-dimensional approach to chiral
  symmetry breaking,''
  JHEP {\bf 0601}, 157 (2006)
  [arXiv:hep-ph/0510268].

  \bibitem{Hirn:2005nr}
  J.~Hirn and V.~Sanz,
   ``Interpolating between low and high energy QCD via a 5D Yang-Mills  model,''
  JHEP {\bf 0512}, 030 (2005)
  [arXiv:hep-ph/0507049];
  J.~Hirn, N.~Rius and V.~Sanz,
  ``Geometric approach to condensates in holographic QCD,''
  Phys.\ Rev.\ D {\bf 73}, 085005 (2006)
  [arXiv:hep-ph/0512240].

  \bibitem{Ghoroku:2005vt}
  K.~Ghoroku, N.~Maru, M.~Tachibana and M.~Yahiro,
  ``Holographic model for hadrons in deformed AdS(5) background,''
  Phys.\ Lett.\ B {\bf 633}, 602 (2006)
  [arXiv:hep-ph/0510334].

  \bibitem{Polchinski:2002jw}
  J.~Polchinski and M.~J.~Strassler,
  ``Deep inelastic scattering and gauge/string duality,''
  JHEP {\bf 0305}, 012 (2003)
  [arXiv:hep-th/0209211].

\bibitem{Brower:2006ea}
  R.~C.~Brower, J.~Polchinski, M.~J.~Strassler and C.~I.~Tan,
  ``The pomeron and gauge / string duality,''
  arXiv:hep-th/0603115.

\bibitem{Babington:2003vm}
  J.~Babington, J.~Erdmenger, N.~J.~Evans, Z.~Guralnik and I.~Kirsch,
  ``Chiral symmetry breaking and pions in non-supersymmetric gauge /  gravity
  Phys.\ Rev.\  D {\bf 69}, 066007 (2004)
  [arXiv:hep-th/0306018].

 \bibitem{Peeters:2007ab}
  For a recent review of the top-bottom approach to the string/gauge correspondence,
 including applications  to finite temperature and phase transitions, see:
  K.~Peeters and M.~Zamaklar,
  ``The string/gauge theory correspondence in QCD,''
  arXiv:0708.1502 [hep-ph] and references therein.


\bibitem{Karch:2006pv}
  A.~Karch, E.~Katz, D.~T.~Son and M.~A.~Stephanov,
  ``Linear confinement and AdS/QCD,''
  Phys.\ Rev.\  D {\bf 74}, 015005 (2006)
  [arXiv:hep-ph/0602229].

\bibitem{Metsaev:1999kb}
  R.~R.~Metsaev,
  ``IIB supergravity and various aspects of light-cone formalism in AdS
  space-time,''
  arXiv:hep-th/0002008.

\bibitem{Brodsky:1997de}
  S.~J.~Brodsky, H.~C.~Pauli and S.~S.~Pinsky,
  ``Quantum chromodynamics and other field theories on the light cone,''
  Phys.\ Rept.\  {\bf 301}, 299 (1998)
  [arXiv:hep-ph/9705477].

\bibitem{Soper:1976jc}
  D.~E.~Soper,
  ``The Parton Model And The Bethe-Salpeter Wave Function,''
  Phys.\ Rev.\  D {\bf 15}, 1141 (1977).

\bibitem{Drell:1969km}
  S.~D.~Drell and T.~M.~Yan,
  ``Connection Of Elastic Electromagnetic Nucleon Form-Factors At Large Q**2
  And Deep Inelastic Structure Functions Near Threshold,''
  Phys.\ Rev.\ Lett.\  {\bf 24}, 181 (1970).

\bibitem{West:1970av}
  G.~B.~West,
  ``Phenomenological model for the electromagnetic structure of the proton,''
  Phys.\ Rev.\ Lett.\  {\bf 24}, 1206 (1970).

\bibitem{Brodsky:1980zm}
  S.~J.~Brodsky and S.~D.~Drell,
  ``The Anomalous Magnetic Moment And Limits On Fermion Substructure,''
  Phys.\ Rev.\  D {\bf 22}, 2236 (1980).

\bibitem{Breitenlohner:1982jf}
  P.~Breitenlohner and D.~Z.~Freedman,
  ``Stability In Gauged Extended Supergravity,''
  Annals Phys.\  {\bf 144}, 249 (1982).


\bibitem{Infeld:1941}
  L. Infeld,
  ``On a New Treatment of Some Eigenvalue Problems'',
  Phys.\ Rev. {\bf 59}, 737 (1941).

  \bibitem{Eidelman:2004wy}
  S.~Eidelman {\it et al.}  [Particle Data Group],
  ``Review of particle physics,''
  Phys.\ Lett.\  B {\bf 592}, 1 (2004).


\bibitem{Brodsky:1974vy}
  S.~J.~Brodsky and G.~R.~Farrar,
  ``Scaling Laws For Large Momentum Transfer Processes,''
  Phys.\ Rev.\  D {\bf 11}, 1309 (1975).

\bibitem{Matveev:1973ra}
  V.~A.~Matveev, R.~M.~Muradian and A.~N.~Tavkhelidze,
  ``Automodellism in the large - angle elastic scattering and structure of
  hadrons,''
  Lett.\ Nuovo Cim.\  {\bf 7}, 719 (1973).

\bibitem{Ferrara:1972qz}
  S.~Ferrara, A.~F.~Grillo and G.~Parisi,
  ``Conformal symmetry at lightlike distances and asymptotic behaviour of
  electromagnetic form factors,''
  Nuovo Cim.\  A {\bf 12}, 952 (1972).

\bibitem{Baldini:1998qn}
  R.~Baldini, S.~Dubnicka, P.~Gauzzi, S.~Pacetti, E.~Pasqualucci and Y.~Srivastava,
  ``Nucleon time-like form factors below the N anti-N threshold,''
  Eur.\ Phys.\ J.\  C {\bf 11}, 709 (1999).

\bibitem{Tadevosyan:2007yd}
  V.~Tadevosyan {\it et al.}  [Jefferson Lab F(pi) Collaboration],
  ``Determination of the pion charge form factor for \mbox{$Q^2$}=0.60-1.60
  \mbox{GeV$^2$},''
  Phys.\ Rev.\  C {\bf 75}, 055205 (2007)
  [arXiv:nucl-ex/0607007].

\bibitem{Horn:2006tm}
  T.~Horn {\it et al.}  [Fpi2 Collaboration],
  ``Determination of the charged pion form factor at Q**2 = 1.60-(GeV/c)**2
  and 2.45-(GeV/c)**2,''
  Phys.\ Rev.\ Lett.\  {\bf 97}, 192001 (2006)
  [arXiv:nucl-ex/0607005].

\bibitem{Pauli:1985ps}
  H.~C.~Pauli and S.~J.~Brodsky,
  ``Discretized Light Cone Quantization: Solution To A Field Theory In One
  Space One Time Dimensions,''
  Phys.\ Rev.\  D {\bf 32}, 2001 (1985).

\bibitem{Radyushkin:2006iz}
  A.~V.~Radyushkin,
  ``Holographic wave functions, meromorphization and counting rules,''
  Phys.\ Lett.\  B {\bf 642}, 459 (2006)
  [arXiv:hep-ph/0605116].

\bibitem{Peskin:1995ev}
  M.~E.~Peskin and D.~V.~Schroeder,
  ``An Introduction To Quantum Field Theory,''
{\it  Reading, USA: Addison-Wesley (1995).}

\bibitem{Choi:2006ha}
  H.~M.~Choi and C.~R.~Ji,
  ``Conformal Symmetry and Pion Form Factor: Soft and Hard Contributions,''
  Phys.\ Rev.\  D {\bf 74}, 093010 (2006)
  [arXiv:hep-ph/0608148].

   \bibitem{Aitala:2000hb}
  E.~M.~Aitala {\it et al.}  [E791 Collaboration],
  ``Direct measurement of the pion valence quark momentum distribution, the
  pion light-cone wave function squared,''
  Phys.\ Rev.\ Lett.\  {\bf 86}, 4768 (2001)
  [arXiv:hep-ex/0010043].

\bibitem{Diehl:2005wq}
  M.~Diehl,
  ``Generalized parton distributions from form factors,''
  Nucl.\ Phys.\ Proc.\ Suppl.\  {\bf 161}, 49 (2006)
  [arXiv:hep-ph/0510221].

\bibitem{Brodsky:1989pv}
  S.~J.~Brodsky and G.~P.~Lepage,
  ``Exclusive Processes In Quantum Chromodynamics,''
  Adv.\ Ser.\ Direct.\ High Energy Phys.\  {\bf 5}, 93 (1989).

\bibitem{Sivers:1975dg}
  D.~W.~Sivers, S.~J.~Brodsky and R.~Blankenbecler,
  ``Large Transverse Momentum Processes,''
  Phys.\ Rept.\  {\bf 23}, 1 (1976).

\bibitem{White:1994tj}
  C.~G.~White {\it et al.},
  ``Comparison of 20 exclusive reactions at large t,''
  Phys.\ Rev.\  D {\bf 49}, 58 (1994).

\bibitem{Holt:1990ze}
  R.~J.~Holt,
  ``Exclusive photonuclear reactions and asymptotic scaling,''
  Phys.\ Rev.\  C {\bf 41}, 2400 (1990).

\bibitem{Bochna:1998ca}
  C.~Bochna {\it et al.}  [E89-012 Collaboration],
  ``Measurements of deuteron photodisintegration up to 4.0-GeV,''
  Phys.\ Rev.\ Lett.\  {\bf 81}, 4576 (1998)
  [arXiv:nucl-ex/9808001].

\bibitem{Rossi:2004qm}
  P.~Rossi {\it et al.}  [CLAS Collaboration],
  ``Onset of asymptotic scaling in deuteron photodisintegration,''
  arXiv:hep-ph/0405207.

\bibitem{Rock:1991jy}
  S.~Rock {\it et al.},
  ``Measurement of elastic electron - neutron scattering and inelastic electron
  - deuteron scattering cross-sections at high momentum transfer,''
  Phys.\ Rev.\  D {\bf 46}, 24 (1992).

\bibitem{Brodsky:1976rz}
  S.~J.~Brodsky and B.~T.~Chertok,
  ``The Asymptotic Form-Factors Of Hadrons And Nuclei And The Continuity Of
  Particle And Nuclear Dynamics,''
  Phys.\ Rev.\  D {\bf 14}, 3003 (1976).

\bibitem{Brodsky:1983vf}
  S.~J.~Brodsky, C.~R.~Ji and G.~P.~Lepage,
  ``Quantum Chromodynamic Predictions For The Deuteron Form-Factor,''
  Phys.\ Rev.\ Lett.\  {\bf 51}, 83 (1983).

\bibitem{Danagoulian:2007gs}
  A.~Danagoulian {\it et al.}  [Hall A Collaboration],
  ``Compton scattering cross section on the proton at high momentum transfer,''
  Phys.\ Rev.\ Lett.\  {\bf 98}, 152001 (2007)
  [arXiv:nucl-ex/0701068].

\bibitem{Diehl:1998kh}
  M.~Diehl, T.~Feldmann, R.~Jakob and P.~Kroll,
  ``Linking parton distributions to form factors and Compton scattering,''
  Eur.\ Phys.\ J.\  C {\bf 8}, 409 (1999)
  [arXiv:hep-ph/9811253].

\bibitem{Chen:2001sm}
  A.~Chen,
  ``Proton Anti-Proton Pair Production In Two-Photon Collisions At Belle,''
{\it Photon 2001, Ascona, Switzerland, 2-7 Sep 2001}

\bibitem{Court:1986dh}
  G.~R.~Court {\it et al.},
  ``Energy Dependence Of Spin Effects In P (Polarized) P (Polarized) $\to$ PP,''
  Phys.\ Rev.\ Lett.\  {\bf 57}, 507 (1986).

\bibitem{Mardor:1998zf}
  I.~Mardor {\it et al.},
  ``Nuclear transparency in large momentum transfer quasielastic scattering,''
  Phys.\ Rev.\ Lett.\  {\bf 81}, 5085 (1998).

\bibitem{Brodsky:1987xw}
  S.~J.~Brodsky and G.~F.~de Teramond,
  ``Spin Correlations, QCD Color Transparency and Heavy Quark Thresholds in
  Proton Proton Scattering,''
  Phys.\ Rev.\ Lett.\  {\bf 60}, 1924 (1988).

\bibitem{Dutta:2004fw}
  D.~Dutta and H.~Gao,
  ``The generalized counting rule and oscillatory scaling,''
  Phys.\ Rev.\  C {\bf 71}, 032201 (2005)
  [arXiv:hep-ph/0411267].


\bibitem{Brodsky:1988xz}
  S.~J.~Brodsky and A.~H.~Mueller,
  ``Using Nuclei to Probe Hadronization in QCD,''
  Phys.\ Lett.\  B {\bf 206}, 685 (1988).

\bibitem{Bertsch:1981py}
  G.~Bertsch, S.~J.~Brodsky, A.~S.~Goldhaber and J.~F.~Gunion,
  ``Diffractive Excitation In QCD,''
  Phys.\ Rev.\ Lett.\  {\bf 47}, 297 (1981).


\bibitem{Aitala:2000hc}
  E.~M.~Aitala {\it et al.}  [E791 Collaboration],
  ``Observation of color-transparency in diffractive dissociation of pions,''
  Phys.\ Rev.\ Lett.\  {\bf 86}, 4773 (2001)
  [arXiv:hep-ex/0010044].

\bibitem{Clasie:2007gq}
  B.~Clasie {\it et al.},
  ``Measurement of Nuclear Transparency for the $A(e,e' pi^+)$ Reaction,''
  arXiv:0707.1481 [nucl-ex].

\bibitem{Airapetian:2002eh}
  A.~Airapetian {\it et al.}  [HERMES Collaboration],
  ``The Q**2-dependence of nuclear transparency for exclusive rho0
  production,''
  Phys.\ Rev.\ Lett.\  {\bf 90}, 052501 (2003)
  [arXiv:hep-ex/0209072].



\end{thebibliography}
\end{document}